\pdfoutput=1

\documentclass[11pt]{article}

\usepackage{acl}

\usepackage{times}
\usepackage{latexsym}
\usepackage{xspace}

\usepackage[T1]{fontenc}

\usepackage[utf8]{inputenc}

\usepackage{microtype}

\usepackage{inconsolata}
\usepackage{url}            
\usepackage{booktabs}       
\usepackage{amsfonts}       
\usepackage{nicefrac}       
\usepackage{microtype}      
\usepackage{amsmath}
\usepackage{amssymb}
\usepackage{enumitem}
\usepackage{graphicx}       
\usepackage{caption}
\usepackage{subcaption}
\usepackage{pifont}
\usepackage{multirow}
\usepackage{textcomp}
\usepackage{listings}
\usepackage{flushend}
\usepackage[hang,flushmargin]{footmisc}

\newcommand{\rz}{Rank\-Zephyr\xspace}
\newcommand{\ciral}{CIRAL\xspace}

\newcommand{\insertcrosslingualscenario}{
    \begin{table*}[t]
    \resizebox{\textwidth}{!}{
    
    \begin{tabular}{lcccccccccc}
    \toprule
    & \multicolumn{2}{c}{\textbf{Source}} & \multicolumn{4}{c}{\textbf{nDCG@20}} & \multicolumn{4}{c}{\textbf{MRR@100}} \\
    & \textbf{Prev.} & \textbf{top-k} & \textbf{ha} & \textbf{so} & \textbf{sw} & \textbf{yo} & \textbf{ha} & \textbf{so} & \textbf{sw} & \textbf{yo} \\
    \midrule
    (1a) BM25-QT & None & $|C|$ & 0.0870 & 0.0824 & 0.1252 & 0.2600 & 0.1942 & 0.1513 & 0.3098 & 0.3914 \\
    (1b) BM25-DT & None & $|C|$ & 0.2142 & 0.2517 & 0.2260 & 0.4169 & 0.4009 & 0.4348 & 0.4313 & 0.5359 \\
    
    \midrule
    \multicolumn{11}{l}{\textit{Cross-lingual Reranking: English queries, passages in African languages}} \\
    (2a) RankGPT\textsubscript{4} & BM25-DT & 100 & \textbf{0.3577} & \textbf{0.3268} & \textbf{0.2991} & \textbf{0.4738} & \textbf{0.7006} & \textbf{0.6038} & \textbf{0.6270} & \textbf{0.6732} \\
    (2b) RankGPT\textsubscript{3.5} & BM25-DT & 100 & 0.2413 & 0.2984 & 0.2497 & 0.4413 & 0.5125 & 0.5360 & 0.5577 & 0.6080 \\
    (2c) \rz & BM25-DT & 100 & 0.2741 & 0.2996 & 0.2881 & 0.4218 & 0.4917 & 0.5397 & 0.5823 & 0.5853 \\

    \midrule
    \multicolumn{11}{l}{\textit{English Reranking: English queries, English passages}} \\
    (3a) RankGPT\textsubscript{4} & BM25-DT & 100 & \textbf{0.3967} & \textbf{0.3812} & \textbf{0.3694} & \textbf{0.5355} & \textbf{0.7042} & 0.6313 & \textbf{0.7058} & \textbf{0.6858} \\
    (3b) RankGPT\textsubscript{3.5} & BM25-DT & 100 & 0.2980 & 0.3189 & 0.3010 & 0.4621 & 0.5702 & 0.5826 & 0.6150 & 0.6582 \\
    (3c) \rz & BM25-DT & 100 & 0.3686 & 0.3622 & 0.3601 & 0.4887 & 0.6431 & \textbf{0.6453} & 0.6995 & 0.6467 \\
    \bottomrule
    \end{tabular}
    }
    \caption{Comparison of Cross-lingual and English reranking results. The cross-lingual scenario uses CIRAL's English queries and African language passages  while English reranking crosses the language barrier with English translations of the passages.}
    \label{tab:cross-reranking}
    \end{table*}
    
}

\newcommand{\insertquerytranslationscenario}{
    \begin{table*}[t]
    \resizebox{0.99\textwidth}{!}{
    
    \begin{tabular}{lcccccccccc}
    \toprule
    & \multicolumn{2}{c}{\textbf{Source}} & \multicolumn{4}{c}{\textbf{nDCG@20}} & \multicolumn{4}{c}{\textbf{MRR@100}} \\
    & \textbf{Prev.} & \textbf{top-k} & \textbf{ha} & \textbf{so} & \textbf{sw} & \textbf{yo} & \textbf{ha} & \textbf{so} & \textbf{sw} & \textbf{yo} \\
    \midrule
    (1) BM25-DT & None & $|C|$ & 0.2142 & 0.2517 & 0.2260 & 0.4169 & 0.4009 & 0.4348 & 0.4313 & 0.5359 \\
    
    \midrule
    \multicolumn{11}{l}{\textit{LLM Query Translations: Queries and passages in African languages}} \\
    (2a) RankGPT\textsubscript{4} & BM25-DT & 100 & 0.3458 & \textbf{0.3487} & \textbf{0.3559} & \textbf{0.4834} & 0.6293 & 0.4253 & \textbf{0.6961} & \textbf{0.6551} \\
    (2b) RankGPT\textsubscript{3.5} & BM25-DT & 100 & 0.2370 & 0.2850 &  0.2741 & 0.4190 & 0.4651 & 0.4937 & 0.5295 & 0.5594 \\

    \midrule
    \multicolumn{11}{l}{\textit{GMT Query Translations: Queries and passages in African languages}} \\
    (3a) RankGPT\textsubscript{4} & BM25-DT & 100 & \textbf{0.3523} & 0.3159 & 0.3012 & 0.4386 & \textbf{0.6800} & \textbf{0.5421} & 0.6149 & 0.5935 \\
    (3b) RankGPT\textsubscript{3.5} & BM25-DT & 100 & 0.2479 & 0.2894 & 0.2692 & 0.4001 & 0.4996 & 0.5005 & 0.5539 & 0.5419 \\
    (3c) \rz & BM25-DT & 100 & 0.2515 & 0.2621 & 0.2497 & 0.3873 & 0.4573 & 0.4644 & 0.5401 & 0.5171 \\

    \bottomrule
    \end{tabular}
    }
    \caption{Reranking in African languages using query translations and passages in the African language. BM25-DT is used as first stage. Query translations are done using the LLMs, and we compare effectiveness with GMT translations.}
    \label{tab:qt-reranking}
    \end{table*}
    
}

\newcommand{\insertmtresults}{

    \begin{table}[t]
    \resizebox{0.92\columnwidth}{!}{%
        \begin{tabular}{lrrrr|c}
        \toprule
        Model & \textbf{ha} & \textbf{so} & \textbf{sw} & \textbf{yo} & \textbf{avg} \\
        \midrule
        RankGPT\textsubscript{4} & 21.8 & 7.4 & 43.8 & 16.0 & 22.3  \\
        RankGPT\textsubscript{3.5} & 7.1 & 1.8 & 42.4 & 6.6 & 14.5 \\
        GMT & 45.3 & 17.9 & 85.9 & 36.7 & 46.5 \\
        \bottomrule
        \end{tabular}%
    }
    \caption{Evaluation of the LLMs query translation quality using the BLEU metric. Scores reported are the average over three (3) translation iterations. LLM translations are evaluated against \ciral's human query translation, and results obtained for Google Machine translate are also reported.}
    \label{mt_results}
    \end{table}
}

\title{Zero-Shot Cross-Lingual Reranking with Large Language\\ Models for Low-Resource Languages}

\author{Mofetoluwa Adeyemi, Akintunde Oladipo, Ronak Pradeep, Jimmy Lin \\[1ex]
        David R. Cheriton School of Computer Science \\
        University of Waterloo \\[1ex]
        \texttt{\{moadeyem, aooladipo, rpradeep, jimmylin\}@uwaterloo.ca}}

\begin{document}
\maketitle
\begin{abstract}
Large language models (LLMs) have shown impressive zero-shot capabilities in various document reranking tasks. 
Despite their successful implementations, there is still a gap in existing literature on their effectiveness in low-resource languages.
To address this gap, we investigate how LLMs function as rerankers in cross-lingual information retrieval (CLIR) systems for African languages.
Our implementation covers English and four African languages (Hausa, Somali, Swahili, and Yoruba) and we examine cross-lingual reranking with queries in English and passages in the African languages.
Additionally,
we analyze and compare the effectiveness of monolingual reranking using both query and document translations.
We also evaluate the effectiveness of LLMs when leveraging their \emph{own} generated translations.
To get a grasp of the effectiveness of multiple LLMs, our study focuses on the proprietary models RankGPT\textsubscript{4} and RankGPT\textsubscript{3.5}, along with the open-source model, \rz.
While reranking remains most effective in English, 
our results reveal that cross-lingual reranking may be competitive with reranking in African languages depending on the multilingual capability of the LLM.
\end{abstract}

\section{Introduction}

Several works have demonstrated the effectiveness of large language models (LLMs) across NLP tasks~\cite{Zhou2022LargeLM, Zhu2023MultilingualMT, Wang2023GPTNERNE}.
For text ranking, researchers have explored the effectiveness of LLMs as retrievers~\cite{Ma2023FineTuningLF}, and as pointwise or listwise rerankers. 
Reranking is cast as text generation so that the models either generate an ordered list \cite{sun2023chatgpt, pradeep2023rankvicuna, ma2023zero} or the ordered list is created by sorting the token probabilities generated~\cite{ma2023zero}.
The large context size of LLMs makes listwise approaches particularly attractive because the model attends to multiple documents and produces a relative ordering. 
\citet{ma2023zero} outperforms zero-shot pointwise approach on three TREC web search datasets using a listwise approach. 
Further, their work showed the potential that listwise reranking by LLMs generalizes across different languages.

In this study, we examine the effectiveness of proprietary and open-source models for listwise reranking in low-resource African languages. 
Our investigation is guided by the following research questions:

\begin{itemize}[leftmargin=*]
    \item How well do LLMs perform as listwise rerankers for low-resource languages?
    \item How effectively do LLMs perform listwise reranking in cross-lingual scenarios compared to monolingual (English or low-resource language) scenarios?
    \item When we leverage translation, is reranking more effective when translation is performed using the same LLM used for zero-shot reranking?
\end{itemize}

\noindent This study aims to answer these questions through an extensive investigation of the effectiveness of RankGPT~\cite{sun2023chatgpt} and RankZephyr~\cite{pradeep2023rankzephyr} in cross-lingual and monolingual retrieval settings. 
We use CIRAL \cite{CiralHfCite}, a cross-lingual information retrieval dataset covering four ($4$) African languages, and construct monolingual retrieval scenarios through either document or query translation. 
The cross-lingual scenarios entail searching with English queries and retrieving passages in the African languages.

Our results show that cross-lingual reranking using these models is consistently more effective than reranking in low-resource languages, underscoring the fact that these LLMs are better tuned to English than low-resource languages. Across all languages, we achieve our best results when reranking entirely in English language using retrieval results obtained by document translation. In this setting, we see up to 7 points improvement in nDCG@20 over cross-lingual reranking using RankGPT\textsubscript{4}, and up to 9 points over reranking in African languages. When reranking in African languages, we gain improvements for RankGPT\textsubscript{4} when we perform query translation using GPT-4 itself. However, for RankGPT\textsubscript{3.5}, we see no significant difference in reranking effectiveness when we translate queries using GPT-3.5.

\section{Background and Related Work}

Given a corpus \( C = \{D_1, D_2, ..., D_n\}\) and a query $q$, information retrieval (IR) systems aim to return the $k$ most relevant documents. Modern IR pipelines typically feature multi-stage architecture in which a first-stage \textit{retriever} returns a list of candidate documents which a \textit{reranker} reorders for improved quality~\cite{Asadi2013EffectivenessefficiencyTF, Nogueira2019MultiStageDR, Zhuang2023BeyondYA}.
While earlier work relied on sparse models such as TF-IDF or BM25 \cite{Robertson2009ThePR} as first-stage retrievers, the improved dense representations of pretrained text encoders such as BERT have encouraged research and adoption of dense retrievers \cite{Karpukhin2020DensePR, Ni2021LargeDE}.

More recently, the effectiveness of Transformer decoder models as components of multi-stage IR systems have been explored in greater depth. Researchers have finetuned GPT-like models in the standard contrastive learning framework \cite{Neelakantan2022TextAC, Muennighoff2022SGPTGS, Zhang2023LanguageMA}, and studied different approaches to reranking using both open-source and proprietary GPT models. 
\citet{sun2023chatgpt} evaluate the effectiveness of OpenAI models on multiple IR benchmarks using query, relevance and permutation generation approaches.
While \citet{Qin2023LargeLM} propose a pairwise approach to ranking with LLMs, \citet{ma2023zero} demonstrate the effectiveness of GPT-3 as a zero-shot listwise reranker and the superiority of listwise over  pointwise approaches.

While these works focus on reranking with LLMs, they only cover two African languages---Swahili \& Yoruba.
For both languages, GPT-3 improves over BM25 significantly but still falls behind supervised reranking baselines.
In this work, we examine the effectiveness of these LLMs as components of IR systems for African languages. 
Specifically, we study the effectiveness of open-source and proprietary LLMs as listwise rerankers for four African languages (Hausa, Somali, Swahili \& Yoruba) in the CIRAL cross-lingual IR test collection~\cite{CiralHfCite}.

Cross-lingual Information Retrieval (CLIR) is a retrieval task in which the queries $q_i$ are in a different language from the documents in the corpus $C$. Popular approaches to CLIR include query translation, document translation, and language-independent representations \cite{Lin2023SimpleYE}.
As the focus of this work is on the effectiveness of LLMs as listwise rerankers, we explore document and query translation approaches in this study.

\section{Method}

\subsection{Listwise Reranking}
In listwise reranking, LLMs compare and attribute relevance over multiple documents in a single prompt. As this approach has been proven to be more effective than pointwise and pairwise reranking \cite{ma2023zero, pradeep2023rankvicuna}, we solely employ listwise reranking in this work. For each query $q$, 
a list of provided documents $D_1, ... ,D_n$ is reranked by the LLM, $n$ being the number of documents at a specific prompt.

\subsection{Prompt Design}
We adopt RankGPT's \citep{sun2023chatgpt} listwise prompt design as modified by \citet{pradeep2023rankvicuna}. The input prompt and generated completion are as follows:

\smallskip
\noindent \textit{Input Prompt:}
\begin{lstlisting}
SYSTEM
You are RankGPT, an intelligent assistant 
that can rank passages based on their relevancy 
to the query.
USER
I will provide you with {num} passages, 
each indicated by number identifier []. 
Rank the passages based on their relevance 
to the query: {query}.
[1] {passage 1}
[2] {passage 2}
...
[num] {passage num}
Search Query: {query}
Rank the {num} passages above based 
on their relevance to the search query. 
The passages should be listed in descending 
order using identifiers. The most relevant 
passages should be listed first. The output 
format should be [] > [], e.g., [1] > [2]. 
Only respond with the ranking results, do not 
say any word or explain.
\end{lstlisting}

\noindent \textit{Model Completion:}

\begin{lstlisting}
[10] > [4] > [5] > [6] ... [12]
\end{lstlisting}

\subsection{LLM Zero-Shot Translations}

Query translation is useful for crossing the language barrier in cross-lingual retrieval and reranking settings. We examine the effectiveness of LLMs in this scenario. For a given LLM, we generate zero-shot translations of queries from English to African languages and implement reranking with the LLM using its translations.
With this approach, we are able to examine the ranking effectiveness of the LLM solely in African languages, 
and look out for the correlation between its translation quality and reranking. 
The prompt design for generating the query translation is as follows:

\noindent \textit{Input Prompt:}

\begin{lstlisting}
Query: {query}
Translate this query to {African language}.
Only return the translation, don't say any 
other word.
\end{lstlisting}

\noindent \textit{Model Completion:}

\begin{lstlisting}
{Translated query}
\end{lstlisting}


\section{Experimental Setup}

\subsection{Models}
We implement zero-shot reranking for African languages on three (3) models. 
These include proprietary reranking LLMs---RankGPT\textsubscript{4} and RankGPT\textsubscript{3.5}, using the \texttt{gpt-4} and \texttt{gpt-3.5-turbo} models respectively from OpenAI's API.
To examine the effectiveness of open-source LLMs, we rerank with \rz ~\cite{pradeep2023rankzephyr}, an open-source reranking LLM obtained by instruction-finetuning Zephyr\textsubscript{$\beta$} \cite{tunstall2023zephyr} to achieve competitive performance with RankGPT models.

\subsection{Test Collection}
Models are evaluated on \ciral~\cite{CiralHfCite}, 
a CLIR test collection consisting of four African languages: Hausa, Somali, Swahili and Yoruba. 
Queries in \ciral~ are natural language factoid questions in English 
while passages are in the respective African languages. 
Each language comprises between 80 and 100 queries, 
and evaluations are done using deep relevance judgements obtained from the passage retrieval task.\footnote{\url{https://ciralproject.github.io/}}
We also make use of \ciral's translated passage collection,\footnote{\url{https://huggingface.co/datasets/CIRAL/ciral-corpus\#translated-dataset}} 
in our document translation use cases. The test collection's documents were translated using the NLLB machine translation model.\footnote{\url{https://huggingface.co/facebook/nllb-200-1.3B}}

We report nDCG@20 scores following the test collection standard,
and MRR@100.

\insertcrosslingualscenario

\subsection{Configurations}
First-stage retrieval is BM25 \cite{Robertson2009ThePR} using the open-source Pyserini \cite{Lin_etal_SIGIR2021_Pyserini} toolkit. 
We use whitespace tokenization for passages in native languages and the default English tokenizer for the translated passages. 
We investigate first-stage retrieval using document (BM25-DT) and query translation (BM25-QT). 
For BM25-QT, we translate queries using Google Machine Translation (GMT). 

We rerank the top 100 passages retrieved by BM25 using the sliding window technique by \citet{sun2023chatgpt} with a window of 20 and a stride of 10. 
We use a context size of 4,096 tokens for RankGPT\textsubscript{3.5} and 8,192 tokens for RankGPT\textsubscript{4}. 
These context sizes are also maintained for the zero-shot LLM translation experiments. 
For each model, translations is done over 3 iterations and we vary the model's temperatures from 0 to 0.6 to allow variation in the translations. Translations are only obtained for the GPT models considering that \rz~is suited only for reranking.

\section{Results}

\subsection{Cross-Lingual vs. Monolingual Reranking}
\autoref{tab:cross-reranking} compares results for the cross-lingual reranking using \ciral's queries and passages as is, and English reranking scenarios.  
Row (1) reports scores for two baselines, BM25 with query translation (BM25-QT) and document translation (BM25-DT).
Cross-lingual reranking scores for the different LLMs are presented in Row (2), and we employ BM25-DT for first-stage retrieval given it is the more effective baseline. 
Scores for reranking in English are reported in Row (3), and results show this to be the more effective scenario across the models and languages.

Improved reranking effectiveness with English translations is expected, given that LLMs, despite being multilingual, are more attuned to English.
The results obtained from reranking solely with African languages further investigate the effectiveness of LLMs in low-resource language scenarios.
We report scores using query translations in \autoref{tab:qt-reranking}, 
with BM25-DT also as the first-stage retriever.
Scores for using the query translations obtained from the specific LLM are reported in Row (2), i.e., 
results in Row (2b) use query translations from RankGPT\textsubscript{3.5} and rerank with RankGPT\textsubscript{3.5}.
The obtained results are a fusion over the $3$ translation iterations using Reciprocal Rank Fusion (RRF) \cite{cormack2009reciprocal}.
In comparing results from the query translation scenario to the cross-lingual results in Row (2) of \autoref{tab:cross-reranking}, we generally observe competitive effectiveness in cross-lingual, and monolingual reranking in the African languages. 
Specifically, RankGPT\textsubscript{4} obtains higher scores for Swahili and Yoruba in the African language scenario, 
especially with its query translations (comparing Rows (2a) in \autoref{tab:cross-reranking} and \ref{tab:qt-reranking}). 

\subsection{LLM Reranking Effectiveness}
We compare the effectiveness of the different LLMs across the reranking scenarios.
RankGPT\textsubscript{4} generally achieves better reranking among the 3 LLMs as presented in the Tables \ref{tab:cross-reranking} and \ref{tab:qt-reranking}.
In the cross-lingual and English reranking scenarios, open-source LLM \rz~\cite{pradeep2023rankzephyr} achieves better reranking scores in comparison with RankGPT\textsubscript{3.5} as reported in Rows (*b) and (*c) in \autoref{tab:cross-reranking}. 
\rz~also achieves comparable scores with RankGPT\textsubscript{4} in the English reranking scenario, and even a higher MRR for Somali as reported in Row (3c) of \autoref{tab:cross-reranking}. 
However, Row (3) in \autoref{tab:qt-reranking} shows that both GPT models achieve better reranking effectiveness compared to \rz~ in the query translation scenario. Comparison with \rz~in the query translations scenario is done only with translations from GMT.
Albeit, these results still establish the growing effectiveness of open-source LLMs for various language tasks considering the limited availability of closed-source LLMs, but with room for improvement in low-resource languages.

\insertquerytranslationscenario
\insertmtresults

\subsection{LLM Translations and Reranking}
Given that RankGPT\textsubscript{4} achieves better reranking effectiveness using its query translations in the monolingual setting, 
we further examine the effectiveness of this scenario.
Row (2) in \autoref{tab:qt-reranking} reports results using LLMs translations, and we compare these to results obtained using translations from GMT. 
Compared to results obtained with GMT translations, RankGPT\textsubscript{4} does achieve better monolingual reranking effectiveness in the African language using its query translations.
There is also a difference in the effectiveness of both translation types when compared with cross-lingual reranking, as using RankGPT\textsubscript{4}'s translations is more effective than the cross-lingual scenario, however, the cross-lingual scenario is generally more effective than using GMT query translations.
RankGPT\textsubscript{3.5} on the other hand achieves less competitive scores using its query translations when compared to translations from the GMT model.

Considering translation quality's effect on reranking, we evaluate the LLMs' translations and report results in \autoref{mt_results}. 
Evaluation is done against \ciral's human query translations using the BLEU\footnote{\url{https://github.com/mjpost/sacrebleu}} metric. 
We observe better translations with RankGPT\textsubscript{4}, and RankGPT\textsubscript{3.5} having less translation quality. Hence, in addition to the model's capabilities, this could be a contributing factor to its reranking results when using its translations. Translations obtained from GMT have the best quality among the three, considering the nature of the model.
Notwithstanding, RankGPT\textsubscript{4} still performs better using its query translations, indicating a correlation in the model's understanding of the African languages.

\section{Conclusion}
In this work, we implement zero-shot cross-lingual reranking with large language models (LLMs) on African languages. Using the list-wise reranking method, our results demonstrate that reranking in English via translation is the most optimal. We examine the effectiveness of the LLMs in reranking for low-resource languages in the cross-lingual and African language monolingual scenarios and find that the LLMs have comparable performances in both scenarios but with better results in cross-lingual. In the process, we also establish that good translations obtained from the LLMs do improve its reranking effectiveness in the African language reranking scenario as discovered with RankGPT\textsubscript{4}.

Our implementation covered three reranking (3) LLMs: RankGPT\textsubscript{4}, RankGPT\textsubscript{3.5} and \rz~and although results indicate RankGPT\textsubscript{4} to be the most effective reranker, they also demonstrate the growing effectiveness of open-source LLMs in reranking for low-resource languages when comparing \rz~and RankGPT\textsubscript{3.5}. 

We believe our work further highlights the capabilities of large language models in tasks regarding low-resourced languages and indicates the prospects that exist for these languages. We additionally hope it encourages research efforts towards the development of methods that improve the effectiveness of LLMs on low-resource languages.

\section*{Acknowledgements}
This research was supported in part by the Natural Sciences and Engineering Research Council (NSERC) of Canada.
\bibliography{custom}




\end{document}